# Forensics Analysis of Xbox One Game Console

Ali M. Al-Haj
*School of Computing*
*University of Portsmouth*
*United Kingdom*

**Abstract**
Games console devices have been designed to be an entertainment system. However, the 8th generation games console have new features that can support criminal activities and investigators need to be aware of them. This paper highlights the forensics value of the Microsoft game console Xbox One, the latest version of their Xbox series. The Xbox One game console provides many features including web browsing, social networking, and chat functionality. From a forensic perspective, all those features will be a place of interest in forensic examinations. However, the available published literature focused on examining the physical hard drive artefacts, which are encrypted and cannot provide deep analysis of the user's usage of the console. In this paper, we carried out an investigation of the Xbox One games console by using two approaches: a physical investigation of the hard drive to identify the valuable file timestamp information and logical examination via the graphical user interface. Furthermore, this paper identifies potential valuable forensic data sources within the Xbox One and provides best practices guidance for collecting data in a forensically sound manner.
*Keywords:* Xbox One, Embedded System, Live Investigation, Games Console, Digital Forensics

**Introduction**

Video game consoles have become a significant part of many peoples' lives. As these devices evolve, it is clear that there is a requirement for digital forensics investigators to identify what they are capable of offering and how it can be examined. It is no secret that these devices challenge investigators in many ways because such security features involve encrypted data. These challenges should not stop investigators. Research is required into how to deal with these devices in an efficient way. The 8th generation of game consoles devices requires connection to the Internet to take advantage of the devices' features. People use games consoles not only to play games but for browsing the web, sharing videos, watching TV and communicating with others – among many other features these devices can offer. The capability of these devices makes it more capable than just playing games. The Xbox One is the 'all in one entertainment system' as Microsoft likes to define it (Xbox, 2015). Xbox One was released with two versions; the first one with a 500 GB hard drive in November 2013 and the second one with a 1TB hard drive in the fall of 2015. Moreover, this device has been designed to be a multimedia device, like a PC, Microsoft has made the Xbox One more capable to the end user and, therefore, more valuable for the digital forensics investigator, compared to other game consoles. Every game played, every web page visited and every message sent can leave digital traces on the user's device or profile. Currently, no one can deny the fact that game console forensic analysis has become such an important and independent part of digital forensic investigations. There are over 13 million Xbox One game consoles in worldwide circulation (Orland, 2015). However, little literary guidance is available offering forensic investigators an insight into what information of interest is stored on the Xbox One, or how to gain information from a forensics perspective. This research aims to provide a best practical guidance to investigate the Xbox One games console that can help investigators deal with such a system. Furthermore, it demonstrates two approaches; a physical investigation into the hard drive, to identify valuable file timestamp information, furthering the work conducted by Moore et al. (2014), and a logical investigation via the graphical user interface (GUI).

The structure of this paper is arranged as follows. Section 2 covers the literature related to the analysis conducted on other 8th generation games consoles and also identify similar forensics challenges, Section 3 describes the initial study conducted which is used to design the experiment methodology, Section 4 describes the experiment methodology, Section 5 states the forensic analysis of the Xbox One, Section 6 describes the additional tests that have been conducted, Section 7 presents the proposed guidance for examination of the Xbox One, and, finally, Section 8 highlights conclusions, and future work.

**Background**

There are two published work on Xbox One forensics, first one was done by (Moore et al., 2014) their research provided the initial analysis of the device and its hard drive. However, in their study they used the traditional experimental methods in digital forensics investigation (Davies et al., 2015) such as file carving, keyword searches, network forensics and file system analysis. According to (Moore et al., 2014) the greatest challenge seems to be the encrypted and/or compressed nature of the hard drive and the network traffic, therefore causing extraction and analysis slightly difficult. However, an analysis of the NTFS file system during this controlled experiment enabled them to retrieve file timestamps. In their findings they stand on some hypotheses for some files functionality these need to be tested. Moreover, they did a live investigation on some encrypted network traffic that could be related back to which game was played. The Xbox One uses a standard file system (NTFS) which provides files timestamp examination unlike PS4, which has a totally encrypted hard drive (Davies et al., 2015). However, according to (Moore et al., 2014) the encrypted and/or compressed nature of the file system, make it hard to provide an in-depth analysis of the Xbox One operating system artefacts. The second research was done by (Shields, 2014 ) the research focused on the encrypted hard drive artefacts, the researcher has focused on examining the NTFS file timestamp modification during the experiment and notified the new files created after each test. However, (Moore et al., 2014) were able to identify the installed games and applications from the User contact partition on the hard drive by identifying the unique 36 characters that related to each game also the in the user partition has the same size of the game or application. Though (Shields, 2014) provided a larger list of the available applications on the Xbox live store with its unique characters that helps the investigator to identify the installed games and applications from the hard drive image.

**Sony PlayStation 4**

The analysis conducted by (Davies et al., 2015) which is also the only published work about the PlayStation 4 forensics they provided the best practical methodology to examine the PlayStation 4 artefacts. Moreover, they developed a special experiment methodology to extract the information from the device after they faced the non-standard file system in the encrypted PlayStation 4 hard drive (Davies et al., 2015). They had no choice but the logical route to investigate the device artefacts. Their study was based on empirical research to identify the features of interest for forensics investigators. The researchers have considered the online and offline investigations by using the Voom Shadow 3 write blocker. Furthermore, the amount of obtainable data from the PlayStation is directly dependent on the operating system version installed on the console (Davies et al., 2015).

**The Forensic Challenges**

The Xbox One uses standard file system (NTFS) that allows file's timestamp examination. However, these timestamp analysis may help investigator to determine if one incident happened on the device or not, such as identifying if there is an installed application on the system or not, it could tell the investigator when was the last shutdown of the device (Moore et al., 2014) but it can never provide a deep investigation to the system artefacts where the information itself is unavailable. However, we cannot ignore the standard NTFS timestamp that is available for us on the hard drive that may provide significant evidence in some cases but it cannot offer the full image. We identified a number of forensics challenges that stood in the way of Xbox One examination in the previous studies. These challenges were shared by Xbox One with most modern embedded system. It starts with the encrypted nature of the hard drive, also the need of the internet contention where most of the user data is stored in the cloud, and finally the data modification through several platforms such as the Smartglass applications.

**The Association of Chief Police Officers (ACPO) Consideration**

Electronic evidence is valuable evidence, and it should be treated in the same manner as traditional forensic evidence with respect and care (Association of Chief Police Officers, 2012). For that, any data extraction methodology presented should be taken into account according of the criteria and standards adopted. This research was carried out in the United Kingdom, and all the examination on the Xbox One implemented with sufficient respect to the current best practice of evidence acquisition provided by Association of Chief Police Officers (ACPO).

**The Initial Study**

The unique Xbox One system specification as we discussed previously in the literature review section, and the empirical research nature, requires initial study and tests. This stage aims to determine the best experiment methodology to meet the digital forensics community needs and aspirations. The initial study started by examining the available literature on Xbox One and others devices such as PlayStation4 and Xbox 360 game consoles to identify the features that the investigators were looking for examining these devices. Furthermore, an empirical investigation through the Xbox One screen menu has been done to navigate the features that can provide evidence of user usage and communication during an examination process.

In this stage, the focus of the investigation was on the features that can provide answers to the digital forensics investigators questions that related to the user identity, the type of information hold time and location. The Xbox One is designed to hold up to 32 account profiles stored on the console according to Xbox Support (2015) these accounts responsible for creating any evidence and can identify the user interaction. Data is generated from many features of the device, the timestamp data created in the hard drive file system or live via the Xbox GUI shows when the data was generated. Finally, the Xbox One system stores the user's data locally on the device in the internal or external storage and the cloud. See Table 1

**The Initial hard drive test:**

The internal Xbox One hard disk was designed by Microsoft to be unchangeable. However, some people succeed to replace the internal hard drive by cloning the original hard disk system partition into a new hard drive using tools box on the UNIX platform with the dd utility. In turn the Xbox system was successfully booted without any issues and without affecting the NAND memory.

**Live Investigation Verifying Methods**

**Write Blockers Tests**

In game console devices, examiners apply write blockers technology to do the live investigation on the games console (Davies et al., 2015) by connecting the write blockers between the game console hard drive and its host device as a man in the middle. However, we tried two write blockers devices on the Xbox One game console to determine our path in the research and to meet our basic forensic integrity.

**The Voom Shadow 3**

The Voom Shadow 3 is a caching hardware write blocker intended to allow a safe, active examination of a system to take place. However, we tried to use it with Xbox One game console to do the live investigation through it, but all ours attempts fail to run the console effectively. Moreover, the researcher has contact the Voom Technology company, and they stated that they have never tried the device with Xbox One and cannot confirm that its work. (See Figure 1).

**Tableau T3458is Forensics Bridge**

The Tableau T3458is Forensics Bridge offers write blocker functionality to the source drive. However, we tested with Xbox One game console as a man in the middle between the Xbox One and its hard drive (See Figure 2).The device was stuck at the initial green screen which cussed because the Xbox one trying to write into the hard drive .

| Feature | Reason |
|---|---|
| Xbox Live Entertainment Network | Most features available to Xbox One users are reliant upon an Xbox live network membership. The device cannot be accessible without an Xbox live membership. However, Xbox One allows the user to add Gusts offline accounts after adding a primary account on the console.<br>The membership enables users to view Xbox live content through a PC web browser, Xbox app on Windows 8, 8.1,10 and smartphones through Xbox One Smartglass app. These accesses will reveal the user's real name, address, credit or debit card information, transaction history, linked devices and sub-account information. |
| The Internet Explorer browser | Internet Explorer on Xbox One offers full web browsing feature for Xbox users. However, The Internet browser does not support office documents contain Recent tabs, Bing search history and favourites pages with numbers of featured pages such as Twitter and Facebook. |
| Media sharing Apps (Upload ,Upload Studio, GmaeDVR, OneDrive, Twitch ) | The Upload enables players to share, via social media, content recorded by the Xbox One Kinect camera, gameplay video, and screenshots. Also, Upload Studio allow users possess the ability to edit screenshots and to record voiceovers and video cover for video commentary.<br>GameDVR the place where the user finds his gameplay recorded videos and screenshots. Players can upload their videos to OneDrive and broadcasting themselves via Twitch. |
| Messages, Voice messages , Party chat | Messages between individuals and multiple users. Xbox parties allow users to group up and chat with friends through Xbox One  up to eight friends |
| Skype App | Enables users to video chat and voice chat with each other while playing games and using apps. Skype communication history. |
| Activity Feed | Recently user's usage of the device and their friends list including recent gameplay, recent achievements, recent new additions to their friends list. |
| Notifications bar, Activity Alert |  Xbox One console notify users of an events, such as when receiving a new message or have an incoming Skype call. Notification history cannot be deleted, and notifications stay in the notification bar for a week or until there are more than 50 (Xbox, 2015).Activity Alert such as added friends and liked posts. |
| Friends list | User's friends and the link on Skype and Twitter. Also, the Real name can be visible depend on the user's privacy setting, meaning that a user's real name can display in all communications. |
| Xbox Storage | Provides system storage information such as disk usage, application saved data, games and apps list and available disk space. |
| User profile and billing information. | Personal data/ Gamertag generated by the user. |

*Table 1 Features of interest for digital forensics investigators*

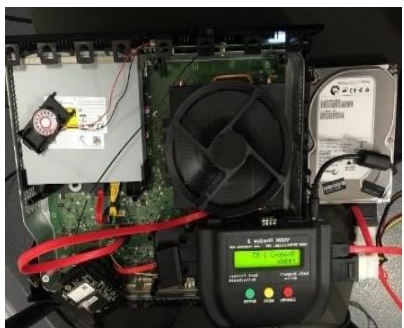

*Figure 1 Voom Shadow 3 connected to Xbox One*

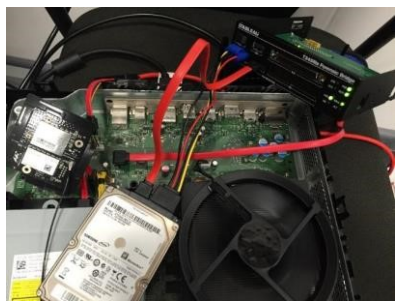

*Figure 2 Tableau T3458is Forensics Bridge connected to Xbox One*

**Proposed Logical Investigation Verifying Method**

Based on our initial findings, and the failing of the tested write blockers technology to work with Xbox One, the best practical way to do live investigation is to forensically clone the original hard drive into new hard drive and lets the cloned image host the examination process. Furthermore, our proposed method keep the original hard drive without modification (ACPO, 2012) (Figure 3) shows the proposed live investigation verifying result method

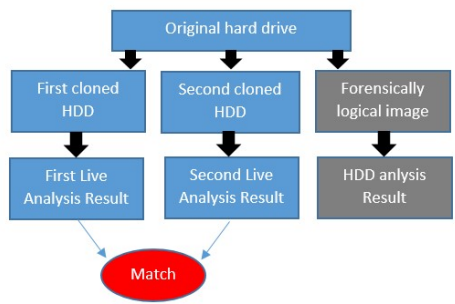

*Figure 3 live investigation verifying method*

**Experiment Methodology**
The proposed methodology for the examination of the Xbox is divided into two analysis approaches: physical and logical, which came after the initial findings presented earlier. Furthermore, the Action phase takes place when the sample of the dataset is introduced into the system via controlled users accounts created for the purpose of examination. Figure 4 shows the process of the experiment. This experiment methodology was developed by studying the features empirically of the interest of forensics examiners that were identified earlier in Table 1.

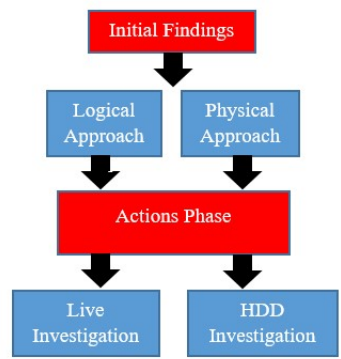

*Figure 4 the Experiment process*

**The Physical approach**
The physical approach aims to identify the hard drive partitions and the timestamp for the existing NTFS files on it and after introducing the set of data. Moreover, it contains some small tests to identify valuable evidence which is provided by the hard drive from the NTFS files timestamp. However, the physical analysis takes into account the previous preliminary forensic analysis of the device that has been done by Moore et al. (2014) to avoid repetition and to build on their findings.

**The Logical Approach**

According to Moore et al. (2014) research findings and the nature of the encrypted hard drive makes the physical analysis ineffective route to provide a deep analysis of the Xbox One system artefacts. Furthermore, the logical analysis route was the primary focus of this research that aims to investigate the Xbox One system via the native Xbox One interface. In addition to that, According to Sutherland et al. (2014) this that route is also the most useful available way to investigate the modern embedded devices.

**The Experiment Methodology Performed on the Device was as followed:**

1. Running video capture camera, recorded time and date to ensure all the examination activities on the system are registered when reference propose is needed.
2. Configured the Xbox One system that needs online access. Recording time and date on the console system to make sure it is compatible with real-time by configuring the right time zone. The device is not accessible to use without this stage.
3. Forensically imaged the Xbox One hard drive for the timestamp offset examination. We used FTK Imager 3.2.2 to create the forensic images.
4. Turned on the Xbox One and recorded the time as set on the console system to note any differences between the real time and the time on the console system. The differences should apply to any data that
   Can be retrieved from Xbox One to ensure that the right time is recorded.
5. Started the Actions phase. Input the Scenarios of Usage. (data inputs)
6. Powered off the Xbox One. Seized the video capture camera and recorded the time. The video capture provides evidence of all data inputs and any changes to the system during this repetition for the set of scenarios.
7. Again, Forensically imaged Xbox One hard drive for the physical analysis stage
8. Forensically cloned the original hard drive image into a new hard drive and left it for validation proposes.
9. Connected the new clone hard disk to the device.
10. Powered on Xbox one investigation areas in (Table 1) to identify activations left by users from the Actions Phase during the experiment loops.
11. Forensically cloned the original hard drive into another hard drive again for validation proposes.
12. Powered on Xbox One investigation areas in (Table 1) to compare with the result from stage number 10.
13. Compared the retrieved data in stages 10, 12, with the input data in the action phase. Recorded information, timestamps and any other evidences indicated by using the device.

**The Actions Phase:**
1. Creating numbers of accounts which are necessary to access Xbox One game console features. Xbox One cannot be used by local accounts. Creating a main Xbox live account on the console device.
2. A Subscription in the Gold Xbox Live membership for an account, to ensure that all the features are available such as party chat. The other users with free membership in Xbox live. However, Microsoft gives all users' permission to access the console system with gold membership to use the features, over 32 users.
3. Adding numbers of contacts and followers to assess the message and other social device functions.
4. Create Skype account to use later for examination. Also adding Skype users to that account to see if they appeared elsewhere on the console.
5. Installing two games (FIFA 2015, Battlefield 4) playing games with different mode (online, offline) and also with single player and multiplayer. However, all the features do not work in the offline playing mode unless the Game DVR recorder.
6. Using a screenshot, Game DVR, Skype calls during play times.
7. Editing recorded video with Kinect camera using Upload Studio application and upload videos to OneDrive.
8. Browsing the internet via Internet Explorer to test recent tabs, pervert tabs, featured tabs and Bing search history.
9. Creating Twitch account and use the live broadcasting feature to see what kind of evidences it can leave.

**Forensic Analysis of Xbox One**
Figure 5 shows a series of different tests that were conducted to consider the ability of a forensic investigator to identify usage of Xbox One. The Action phase was introduced to three operating system versions 6.2.13194.0, 6.2.13326.0 and 6.2.13332. Any notable changes between revisions are discussed below.

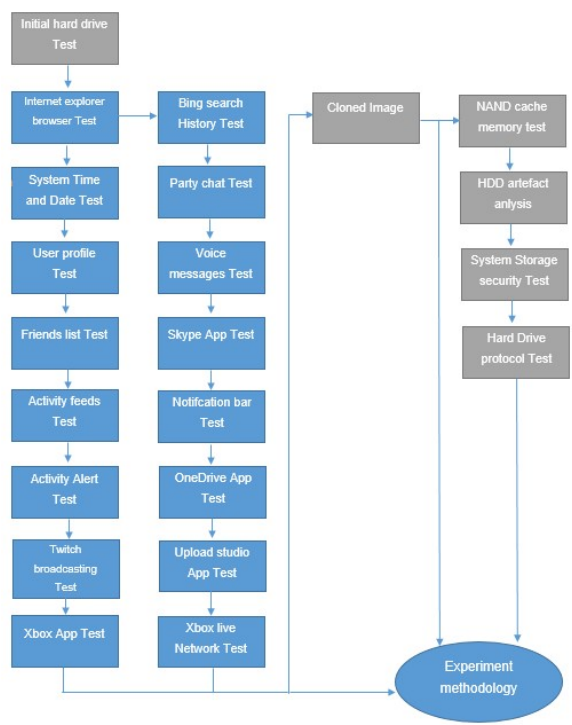

*Figure 5 Empirical examination of the Xbox One*

## The Logical Analysis

### Time and Date Test

We conducted a system-wide analysis of the Microsoft Xbox One, focusing on the retrieval of date and timestamp information. It was discovered that the majority of the features, such as Activity Feed, Activity Alert, etc., provided the data. In contrast, applications such as the Internet Explorer browser did not present any form of a date or time information for the recent tabs. However, Bing search history presents the date, while the Messages features introduced the date and time upon which messages were sent and received. Skype, provided time and date for all retrievable data, OneDrive does not present any data or time for uploaded items.

**Internet Explorer Browser, Favourites Pages, Bing Search History**

A number of tests were carried out on the Xbox One Internet Explorer browser, the experiment involved visiting 106 website. After the consultation with the Xbox support team about the limitation for the recent tabs that are stored in the browser history. However, all the visited web pages has have been recorded in the recent taps tabs.   Bing search history also appears in the browser, as the browser records without logging into the users account in Bing.  The Analysis of the web browser history, favourite pages and featured pages, concluded that the time and date upon visited pages are not available. On the other hand, Bing search history proved date of the search and whether the user has looked for videos, images or general web searches.

Another experiment carried out on the browser to determine if the user leaves  the browser by using the Home button on the Xbox One controller without closing the tab/s whether  it will recover the session or not. During the experiment loop, we discovered that the last session would stay in the browser even if the user logouts from the Xbox One and comes back again. Moreover, this covers the normal browsing and the inPrivate browsing. Furthermore, we tested two default featured pages in the browser, Facebook and Twitter to see whether the browser saves the credential login data or not. However, if the user chooses to save them, they will be retrievable.

**Online Live Results Analysis**

The online investigation on the Xbox One game console was carried out on this research to examine the ability of retrieving valuable information from the system that has been identified earlier in Table 1. However, after several experiments on the three OS versions, we successfully extracted the majority of the data that was identified as a target at the beginning of this research. Unlike PlayStation 4 during our experiment. The firmware version does not affect the amount of the retrievable data from Xbox One. However, the amount of obtainable information is directly attached to the time of the incident and when the investigation process takes place.

Table 2 demonstrates that for each feature.  Table 3 shows the features and types of possible data that could be found.

**Offline Live Results Analysis**

Limited data has been retrieved from the system in the offline mode. However, from the login menu examiner can identify the user/s  has been logged in the device with their firs name and Gamertags, general notification from the notification bar, the game and application available , storage usage and the saved game data. After login. Examiner can browse the gameplay videos recorded in the system.

| Data | Up to 7 days | Up to 30 days | After 30 days |
|---|---|---|---|
| Internet Explorer browser | ✓ | ✓ | ✓ |
| Messages | ✓ | ✓ | X |
| Notifications bar | ✓ | X | X |
| Activity Feed | ✓ | ✓ | X |
| Activity Alert | ✓ | ✓ | X |
| Friends | ✓ | ✓ | ✓ |
| Xbox One Storage | ✓ | ✓ | ✓ |
| Upload Studio (saved video ) | ✓ | ✓ | ✓ |
| Upload Studio (unsaved  video) | ✓ | ✓ | X |
| User profile | ✓ | ✓ | ✓ |
| Skype | ✓ | ✓ | ✓ |
| OneDrive | ✓ | ✓ | ✓ |

✓ = Retrievable     X = Not Retrievable

*Table 2 Xbox One artefact volatility.*

**The Physical Analysis**

**Users Gameplay videos**

Forensically, these recorded game play videos may not have much value for itself as they only show in game footage. However, the user has the ability  of voiceover, pictures or live cam by using the Upload Studio application that allows the contact to be modified. Moreover, Upload Studio allows users to record clips up to 30 minutes with text editing, voice cover, live cam and footage.  Moreover, most of the published crimes are related to Xbox One that have a relation to Kinect camera which can record videos, take pictures and record voices. We examined all these features, when the user records game play clips or take screenshots. The system stores directly on the hard drive and encrypts it with the *.xvd* Microsoft file format. Figure 6 shows three recorded videos we recorded while playing FIFA 2015 and the timestamp of these videos. However, we examined to download these videos used by the Xbox App in Windows 10 and we were able to retrieve the Exit data for these files. Figure 7 shows the encoding time for the video. Moreover, we modified one video that used Upload Studio with voice cover and live cam in 14 /9/2015 17:49 UTC and after downloading the edited video we retrieved the Exit data shown in Figure 8.These findings may help investigators to identify the data and time of the edited videos.

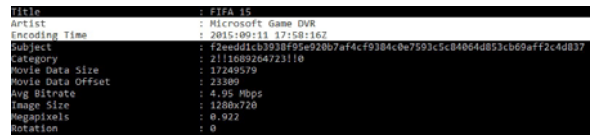

*Figure 7 Exif data for the third recorded videos in Figure 6*

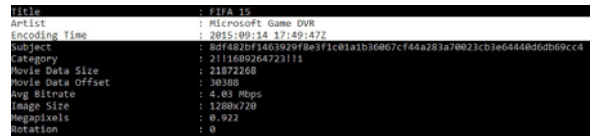

Figure 8 Exif data for the second video in Figure 6 after editing

### Installed Games and Apps Timestamp

We carried out several tests on the games and application timestamp. Moreover, we found that the usage of these games and apps does not affect the modified time on the games and apps at the hard drive in the user partition. This contradicts with (Moore et al., 2014) hypothesised that its show the last time the game was played, and it contain the user playing saved data. We played Battlefield 4 in the offline mode to reach achievement, and the timestamp does not affect by our playing. Moreover, the live offline analysis shows that each player has separately saved data on the hard drive see Figure 9. However, the file timestamp changed when the game was updated and the same for apps. See Figure 10. Table 4 shows the most valuable Applications unique hexadecimal names and size that will help investigators to identify these applications from the hard drive image.

| Xbox One's artefacts | Types of evidences retrievable |
|---|---|
| The Internet Explorer browser | Bing search history, Recent tabs, Website saved passwords, last session, InPrivate last session |
| Messages | Text messages, Voice messages, sharing videos and screenshots |
| Notification bar | Installed items, Party chat invitation, recorded videos, recorded screenshots |
| Activity Feed wall | Posts, commands, Broadcasting time and duration, edited clips, recorded clips, screenshots, how many times games played, adding friends, friends list, friends items, followers items, last played game |
| Login Menu | User/s has been signed in on the device with their Gamertags and first name |
| Activity Alert | Added friends, liked items |
| Friends | Friends, flowers, favourite |
| Xbox One storage | Games and Apps installed, storage space, user's game data. |
| Upload | My captures (All user videos ) |
| User Profile | User first and last name, Gamertags |
| System | Xbox name |
| Billing information | Payable Account, Credit cards, billing address |
| Subscription | Type of membership expired date for gold membership |
| Skype | Calls history, messages, contacts |
| OneDrive | Edited videos, screenshots, external media |

Table 3 Features and types of possible data could be found in it.

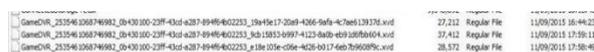

*Figure 6 Recorded videos located on Temp Partition*

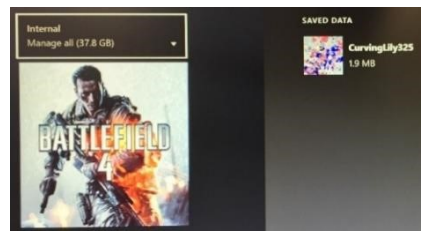

Figure 9 Batlefild4 Saved data with the Gamertag from the live investigation

*Figure 10 Games and Apps located in the User partition*

### Last Session duration

During the repetition of the experiments, we noted that the *.xvi* files situated on the system support partition that got the same 32 name for the applications and game modified each time the device start. Furthermore, it has been known from (Moore et al., 2014) finding and our observation that the *DataCollictionUpdater_0* located in the same partition modified each time the system shut down. To collect last session duration of device we identified the final shutdown of the system and the last run of it. However, the user should have installed at least one application or game before last session to be able to identify the previous session duration. Figure 11 shows that the last session duration was on our system 15 minutes long.

| App Name | hexadecimal name in User contacts partition | Original size | *Compressed size |
|---|---|---|---|
| Skype | 242BF9CE-DA7C-4872-805E-E873ADB32C07 | 54.9 MB | 56,912 MB |
| OneDrive | 26F1625C-A8A5-4481-B6BF-C334F518F845 | 10.8 MB | 11,104 MB |
| Upload Studio | 265C6BAD-C47C-4A63-A1925818037FAC97 | 259.2 MB | 265,444 MB |
| Vine | 1333BF8D-5D7E-41BB-B508 13BCDFF25F43 | 24.8 MB | 25,460 MB |
| YouTube | 13096BD0-8237-47FA-80BE-29A3563CF0BF | 56.9 MB | 58,280 MB |
| Twitch | D0134385-33C0-4382-BE31-58C4CF4F453E | 24.1 MB | 25,460 MB |
| Movies & TV | B0655109-C128-4519-9E36-0D370809CD0E | 48.2 MB | 55,552 MB |
| Media Player | 05D64F3E-9E29-47CE-A23C-1E86F2AFB09A | 45.5 MB | 46,660 MB |
| Netflix | FB4C8FF5-ED19-48FEA462-851A076663C0 | 58.2 MB | 59,660 MB |

*Table 4 Applications name in User content partition*
*Applications size with encrypted and/or compressed data

*Figure 11 System Support Partition*

## Additional Tests

### Kinect Facial Recognition Login

One of the precious points that (Davies et al., 2015) mentioned in their future considerations for the PlayStation 4 that to test how smart the Kinect Facial recognition in the games console. Moreover, we carried out a several experiment on the Xbox One Kinect to fool the camera using big banner, toy hands and eyes move. However, all our attempts failed, which show how this Kinect sensitive and powerful, this finding may use to prove account ownership on a multi-user system if an individual was able to unlock a particular account available on the login menu.

### Protocol Analyser Test

We employed the SATA Sierra M6-1 SATA protocol system to test the traffic between the hard drive and the Xbox One device. However, all the traffic between the device and its hard desk seemed to be encrypted. However, extract some plain text was viewed the hard disk *serial number and model*. (See Figure 12)

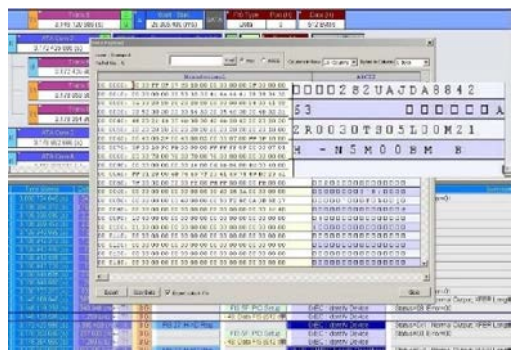

*Figure 12 Hard drive S/N as displayed on the SATA Sierra M6-1 SATA protocol system*

### System Storage Security

We carried out a number of tests on the Xbox One internal hard drive, and we found that it accept modification in the system partition, from forensics perspective that mean it could be used to hide data. However, we stored Alternate Data Streams (ADS) and bootable Virtual Machine (VM) on the hard drive in the five Xbox One system partitions, and the system booted up normally. We noticed that if you store any data on the user partition, it will appear during the live investigation in the storage. Moreover, the user partition size in the Xbox game console, 365 GB in the first version of the Xbox One that released with 500GB hard drive and 780.8 GB in the new version that released with 1TB hard drive.

### Proposed best practice guidance for forensics investigation of Xbox One game console: The Operating System Version

In Xbox One, the Operating system version should match on the hard drive and the NAND memory. Moreover, Xbox One should not force the Examiner to update the operating system if it is on the manually update mode. We recommend the examiner to identify the operating system version on the device before start the online investigation. He can do that, from the hard drive image by the date of the last update from the update partition on the hard disk and using the Microsoft OS versions released date list on their website. The version installed on the system in Figure 6 is 6.2.13194.0 released in 7/7/2015 (see figure 2)

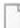

*Figure 12 Updater.xvd from System Update partition last update*

OS version: 6.2.13194.0 (xb_rel_1507.150702-2154) fre

Release date

7/7/2015

*Figure 13 OS Version and release data (Xbox, 2015)*

However, if an investigator faces a forced update should not update the system and keep the investigation offline on the suspect device. Moreover, we suggest that if the Xbox One live login credential available live investigation could take place on another Xbox One console and all the data will be available unless installed games and the notification bar data. Though, the online investigation can also take a place on Xbox One Application in Windows 10 using the User login credential that offer some of the data identified in Table 3. We suggest that police have to seize the suspect personal PC and personal smartphone to surround the suspect activities around the case. However, If the Xbox One live login credential is available, we tested that changing the suspect account's password cannot prevent the suspect to modify some data if he already logged into his account from one of the Xbox applications its doesn't log him out. In this guidance stages ordered from the volatile to the less volatile. It is important to start with right examination approach analysts should determine what type of data they are looking for based on their cases and using Table 2 and 3.

**Live/Active Investigation:**

1. Remove and forensically image the Xbox One Hard Drive.
2. Forensically clone the Xbox One hard drive into new hard drive.
3. Connect the new cloned hard drive into Xbox One.

**Offline Mode**
4. Active the video capture device and record real time and date.
5. Turn on Xbox One and check time and date on the system to apply any offset on the data retrieved at the end.
6. Navigate to the My Game & App menu identify installed games and Apps.
7. Identify the user/s first name/s and their Gamertags form login menu.
8. Login to user account would allow you to see the user's recorded clips and screenshots (from games not edited videos) if the user was recorded in the same system during playing.
9. Check the notifications bar.
10. Disable the Automatic update if it is enabled from the system setting and record that change if you make it.

**Online Mode**
11. Take the system online and start navigation the evidence as follow:
    a. Check the Notifications bar again.
    b. Activity Feed list
    c. Activity Alert
    d. Messages
    e. User Information
    f. Check Twitter account if the user has registered on the console setting screen.
    g. Internet Browser History: the examiner should be aware that when navigating the Bing search history he will create a new visited page in recently visited pages list. It should record it to avoid changing on the suspect evidence. However, he could use the InPrivate browser to prevent that. Check the featured pages (Facebook, Twitter, MSN) provide it by Internet Explorer in the browser it may contain login credentials information could help you with your case.
    h. Upload Studio/ My captures
    i. Skype
    j. OneDrive: we recommend if the user's login credentials available to test the OneDrive on the web that provide you the uploaded items timestamp.
    k. Friends list
12. Turn off Xbox One
13. Turn off capture video record
14. You could clone the original hard drive again into new hard drive and repeat the live investigation for verification purpose.

**Physical investigation HDD:**

1. Back to the hard disk image you have taken on the being start your physical analyses to gather NTFS files timestamp data and to check the system partitions for any strange files. It is up to the examiner to decide using any advanced forensics investigation tools such as FTK and Encase which has carving functionality that could help if the internal hard drive has modified with external data.

2. Identify when the last session of the Xbox One was user should have installed at less one app to be able to identify the previous session.(See Figure 10)
3. Identify valuable installed application/s on the system from the temp contacts partition using table 3. (See Figure 9)
4. Identify user-recorded videos in playing from the temp contact partition which .(See Figure 6)
5. Identify how many connected controllers has been used in the device from the system partition. (See Figure 11)
6. Determine, when the first usage of the device from the system update partition. *System volume information folder* timestamp.

## Conclusion

The proposed best practice guidance would let digital forensics examiner perform a logical analysis of a write protected Xbox one. Moreover, the physical analysis allows them to gain the valuable NTFS file timestamp from Xbox one hard drive. The modification of information is prevented on the original evidence source, and thus evidential integrity is maintained. The amount of data retrievable however is directly dependent on the incident time and when the investigation takes a place. Table 2 determines that. The live offline analysis provided a little amount of data such as the user's first name, recorded gameplay videos and storage state. However, the live online analysis provides the majority of the user data. Table 3 details the features, and the possible data the investigator could find there. The forced operating system update could interrupt the examiner investigation if the credential login were unavailable. However, it is the best practical guidance to examine the device. The Additional tests provided some useful finding for investigators about the Xbox One hard drive also it is aware them that Xbox One hard drive can be used in steganography cases.

## Future Work

The advanced technology that keeps on developing for users and the interactivity options for the Xbox One that may require investigation in their right. Newly, Microsoft allows Xbox one users to stream their Xbox One games to a Windows 10 PC or tablet, this connection would be interesting to examine. Future research should consider the implication of the Xbox One connection capabilities with the SmartGlass App on smartphones and tablets. Any evidence of the user usage and communications will be a place of interest to investigators.

## Acknowledgements

The author would like to thank the Information Security Research Group (ISRG) at the University of South Wales for the various supports given in completing this work. Special thanks go to Gareth Davies, Konstantinos Xynos and Huw Read.